\documentclass[conference,compsoc]{IEEEtran}

\usepackage{cite}
\usepackage{amsmath,amssymb,amsfonts}
\usepackage{booktabs}
\usepackage{array}
\usepackage{graphicx}
\usepackage{textcomp}
\usepackage{xcolor}
\usepackage{url}
\makeatletter
\g@addto@macro{\UrlBreaks}{\do\/\do-\do\_\do\.\do\?\do\&\do\=}
\makeatother
\usepackage{fvextra}
\usepackage{enumitem}
\usepackage[hidelinks]{hyperref}

\graphicspath{{figs/}{graphs/}}

\def\BibTeX{{\rm B\kern-.05em{\sc i\kern-.025em b}\kern-.08em
    T\kern-.1667em\lower.7ex\hbox{E}\kern-.125emX}}

\begin{document}

\title{Engineered Persuasion: Evaluating Personalized Pretexts in LLM-Generated Spear Phishing}

\author{
\IEEEauthorblockN{Jerson Francia, Derek Hansen, Benjamin Schooley, and Shydra Valynn Murray}
\IEEEauthorblockA{\textit{Department of Electrical and Computer Engineering} \\
\textit{Brigham Young University} \\
Provo, UT 84602, USA \\
jersno@byu.edu (J.F.); dlhansen@byu.edu (D.H.); \\
ben\_schooley@byu.edu (B.S.); shywilli@byu.edu (S.V.M.)}
}

\maketitle

\begin{abstract}
Large language models can insert workplace details into phishing pretexts at low cost, but those details may either support or undermine a message's credibility. We recruited 180 U.S. working adults to evaluate simulated, AI-generated phishing emails in a disclosed survey. The emails used four cumulative levels of information: workplace (Level~1); recipient name and job title; job responsibilities; and coworker/shared-project context (Level~4). Participants rated each message's convincingness from 0 to 100, chose one stated action (open the link, investigate, delete, or report), and explained why their highest- and lowest-rated messages stood out. Across 1,436 valid evaluations, convincingness increased by 2.40 points per personalization level in a sensitivity analysis, while the odds of expressing click intention increased by 28\% per level. Among participants who did not express an intention to click, investigation remained common, reporting declined, and deletion increased. A post-hoc descriptive analysis found higher ratings and click intention for messages from a named person who referenced a supplied coworker than for messages from a department or entity. Qualitative coding showed why added detail could help or hurt: details that matched participants' roles and routines supported credibility, while incorrect, vague, or channel-inappropriate details raised suspicion. Together, the results highlight that personalization is not simply a matter of adding more details: it depends on whether the pretext fits the recipient's work context. We discuss how this distinction can inform workplace cybersecurity training.
\end{abstract}


\section{Introduction}

\label{sec:introduction}
Phishing remains one of the main ways attackers gain initial access, even as organizations improve their technical defenses. The 2026 Verizon Data Breach Investigations Report shows that although cyber breaches are increasingly influenced by vulnerability exploitation, ransomware, and AI-supported activity, attacks targeting people remain a major concern \cite{verizon_dbir_2026}. Reports from the Federal Bureau of Investigation (FBI) Internet Crime Complaint Center (IC3) also show that phishing and spoofing make up a large share of complaints, with business email compromise (BEC) leading to the highest median losses \cite{FBI_IC3_2024}. The Anti-Phishing Working Group (APWG) recorded 1,003,924 phishing attacks in the first quarter of 2025 alone, showing that phishing also remains a high-volume threat \cite{apwg_trends_2025}.

Although malicious actors succeed with both simple and highly sophisticated phishing attacks, users should not be framed as "the weakest link." Susceptibility to phishing is often shaped by workload, time pressure, workplace norms, and imperfect defenses that exist in the organization \cite{zimmermann_renaud_2019}. The continued scale of phishing suggests that attackers continue to find value in making malicious requests appear routine, relevant, and socially appropriate.

Many phishing messages work by presenting a \emph{pretext}: a fabricated scenario or backstory that gives the recipient a reason to treat the message as legitimate \cite{workman_wisecrackers_2008,rajivan_creative_2018}. Personal details can strengthen a pretext when the message uses details that match the recipient's workplace, job, routine, or coworkers \cite{xu_personalized_2023,distler_context_2023}. However, those same details can also make a message less convincing when they are wrong, vague, or inconsistent with normal workplace procedures \cite{distler_context_2023}. Pretexts also depend on who the message claims to be from; a message from a department/entity can carry different authority and familiarity cues compared to a person who references someone you know \cite{Williams2018IJHCS,Steves2020PhishScale}. Because pretexts depend on selecting and arranging contextual details, tools that can quickly draft many plausible variants may change the scale at which attackers can tailor them.

Large language models (LLMs) can reduce the time, expertise, and cost needed to draft convincing phishing content \cite{heiding_devising_2023,khan_offensive_2021,czybik_personalized_2026}. Recent peer-reviewed work therefore treats generative AI as part of a broader social-engineering risk, where models can support message drafting, scaling, and repeated adaptation \cite{schmitt_digital_2024,roy_chatbots_2024}. However, lower generation cost does not mean that every generated phishing message will be as convincing. An LLM-generated pretext still depends on whether the message fits the recipient's work context, uses plausible relationships, and avoids details that expose the scenario as wrong or unusual. For that reason, we study which added workplace details make LLM-generated phishing messages seem more believable, and which details make them seem suspicious.

We therefore examine how the amount and type of workplace detail used in AI-generated spear phishing pretexts shape perceived convincingness and stated action in a controlled survey. Participants knowingly evaluated simulated phishing messages; the study was not designed to test detection in a mixed inbox or estimate operational phishing success. This disclosed design allowed repeated comparisons and collection of participants' reasoning without delivering personalized deception to real inboxes. We use \textit{personalization level} as our measure of personalized pretext depth, meaning how much work-context information was made available for the message. We also analyze coded \textit{impersonation style} (sender/pretext configuration) as a post-hoc exploratory component describing the apparent sender and whether the message invoked a supplied coworker.

Building on mixed findings about LLM phishing persuasiveness \cite{heiding_devising_2023,czybik_personalized_2026}, we compare how participants judged messages generated with four cumulative levels of workplace information. Participants rated convincingness and selected one stated response: open the link, investigate, delete, or report. We use \emph{click intention} to mean selection of the open-link option in this hypothetical survey; it is not observed clicking. The planned comparison concerns personalization depth; associations with sender/pretext configuration are exploratory because the model, rather than the experimental design, selected the sender role.

This paper makes three contributions. First, it provides a within-participant comparison of four cumulative workplace-information tiers rather than treating personalization as simply present or absent. Second, it shows how convincingness and forced-choice stated actions changed across those tiers under the same disclosed survey context, without treating those responses as operational click-through. Third, participant explanations show that personalization depends on contextual fit: accurate role, relationship, and project details can support credibility, while incorrect or channel-inappropriate details can expose the pretext. We also report a clearly secondary analysis of the sender/pretext configurations produced by the model.

\section{Review of Related Literature}

We review research on how people judge spear phishing messages, how personalization and persuasion shape those judgments, and how LLMs facilitate the creation of targeted phishing pretexts.

Recent complaint and incident data show that spear phishing attacks remain a persistent problem despite advances in technical safeguards. IC3 reports show that phishing/spoofing account for the largest share of complaints, while business email compromise (BEC) drives the highest median losses \cite{FBI_IC3_2024}. APWG reports also point to historically high phishing volumes in early 2025 \cite{apwg_trends_2025}. Because different industry reports measure phishing in different ways (e.g. complaints reported to the FBI vs confirmed security incidents and data breaches), we cannot reliably determine that phishing attacks are becoming more or less targeted over time. Instead, we view targeting as a choice attackers make by balancing the cost of tailoring a message and its expected return \cite{verizon_dbir_2026,desai_threatlabz_2025}. How effective this targeting is ultimately depends on how targets interpret and respond to the message.

People's susceptibility to phishing is influenced by the cues present in the message, the recipient's own knowledge and habits, and the setting in which they receive it \cite{benenson_unpacking_2017,oliveira_empirical_2019,alsharida_systematic_2023,lin_susceptibility_2019}. In those messages, easily spoofed surface cues, like padlocks and URL warnings, can create a false sense of legitimacy \cite{Dhamija2006WhyPhishingWorks}; however, training tools can improve awareness to these cues and help users detect phishing \cite{canova_nophish_2015}. Organization-oriented phishing research likewise emphasizes that susceptibility and response depend on workplace context, organizational controls, and human factors together \cite{althobaiti_review_2024}. However, recent research suggest that people can still fall for both crude and polished attacks, so phishing risk is not limited to advanced spear phishing messages \cite{burda_cognition_2024,distler_context_2023}. Thus we focus on one part of that larger picture: how personalized details help build a believable phishing pretext.

We define a phishing pretext as the story behind the message: the scenario, role, or request that makes the message appear legitimate. Personalization is when we include specific details that connect this story to the target (e.g., name, job title, responsibilities or coworker), often gathered from open-source intelligence (OSINT) or internal directories. We define personalization depth as the level of the recipient-specific information made available to the message source. This approach builds on prior work comparing phishing messages created with different amounts of target information, although our four levels are specific to the present study \cite{xu_personalized_2023}. Personalization depth is distinct from persuasion strategy: it describes the information available to create the pretext, whereas persuasion strategies describe how the message encourages action. Strategies such as authority, urgency, social proof, and reward---sometimes called weapons of influence---may appear alone or in combination at any personalization level \cite{oliveira_empirical_2019,ferreira_persuasion_2015,rajivan_creative_2018}. Whether these strategies seem plausible also depends on workplace norms. The same request may seem routine when it comes from an expected person in an appropriate tone, but suspicious when the sender or wording does not fit the relationship \cite{holmes_politeness_2015}.

Sender identity is another part of a pretext, but it must be interpreted together with the message body and the workplace relationship. An unfamiliar sender may seem plausible when the role and request fit. Conversely, a familiar name may seem suspicious when the request or communication channel does not. Prior research finds that authority, role, and premise alignment shape responses in workplace phishing tests \cite{Williams2018IJHCS,Steves2020PhishScale}. Law-enforcement materials on BEC similarly describe executive and vendor impersonation as common tactics that exploit organizational roles and familiar relationships \cite{FBI_BEC,CISA_TA15_BEC}.

Prior lab and field studies generally find that including personal details can raise susceptibility to phishing, although these effects vary by population, setting, and message \cite{jagatic2007social,chiew2018survey,xu_personalized_2023}. Much prior research treats personalization as either present or absent, or compares broad low- and high-information conditions \cite{zhuo2023,alsharida_systematic_2023,xu_personalized_2023}. In the \textit{SpearSim} study, message creators who received more information about a target deceived more recipients than creators with minimal information, and produced more contextually meaningful narratives \cite{xu_personalized_2023}. These findings motivate a closer examination of information depth, but also raise a broader question: when does additional target information make a phishing pretext seem more convincing, and when does it instead undermine credibility?

LLMs add a new reason to revisit this question. They can lower the time, cost, and writing skill needed to create convincing phishing pretexts \cite{khan_offensive_2021,heiding_devising_2023,czybik_personalized_2026}. A NeurIPS workshop demonstration showed that automated messages based on target information could support social engineering campaigns at scale \cite{seymour_generative_2018}. More recent work places generative AI within a broader deception workflow: models can help with drafting messages, generating scenarios, and iterating automatically, while safeguards and proposed mitigations remain insufficient \cite{schmitt_digital_2024,roy_chatbots_2024}. At the same time, research on LLM-assisted lateral phishing shows that the effectiveness of these capabilities depends on the organizational context in which they are deployed \cite{bethany_lateral_2025,weinz2025impact}.

The closest operational comparison to this paper is Czybik et al.'s large-scale field experiment. It used information from web searches to generate LLM spear phishing emails, and measured clicks after delivering them to around 7,700 recipients \cite{czybik_personalized_2026}. That study addresses operational effectiveness and scalability. Our study asks a different question: in a controlled setting, how do people judge messages generated with cumulative tiers of workplace information, what action do they say they would take, and which contextual matches (or errors) explain those judgments? This setting may not estimate inbox behavior, but its repeated ratings and open-ended explanations provide evidence about perceived convincingness, stated action, and context fit across information-depth conditions. Laboratory and survey measures can provide useful comparisons when researchers clearly separate them from field behavior. For example, the Phishing Email Suspicion Test compares laboratory suspicion ratings with field-email outcomes but treats them as distinct measures \cite{hakim_phishing_2021}. We are not aware of prior human-subject work that deals with cumulative information tiers in LLM-generated messages with both quantitative and qualitative analyses.

Together, the literature suggests that perceived legitimacy of spear phishing messages depends on several factors: the information in a pretext, its persuasion strategies, the apparent sender and relationships, and whether the request fits workplace norms. Our analysis focuses on cumulative personalization depth. We separately examine coded impersonation style (sender/pretext configuration) as an exploratory factor, and use participant explanations to identify the characteristics they regarded as convincing or suspicious.

\section{Research Questions}

We posit two research questions and one exploratory question:
\begin{itemize}[leftmargin=3em]
   \item [{\textbf{RQ1.}}] How does personalization level relate to self-reported convincingness and stated action in AI-generated spear phishing pretexts?
   \item [{\textbf{EQ1.}}] What associations appear between impersonation style (sender/pretext configuration) and these outcomes?
   \item [{\textbf{RQ2.}}] Which personalized pretext cues do participants report as making spear phishing messages persuasive or suspicious?
\end{itemize}

\section{Methodology}
\label{sec:method}

\begin{figure*}[t]
  \centering
  \includegraphics[width=\textwidth]{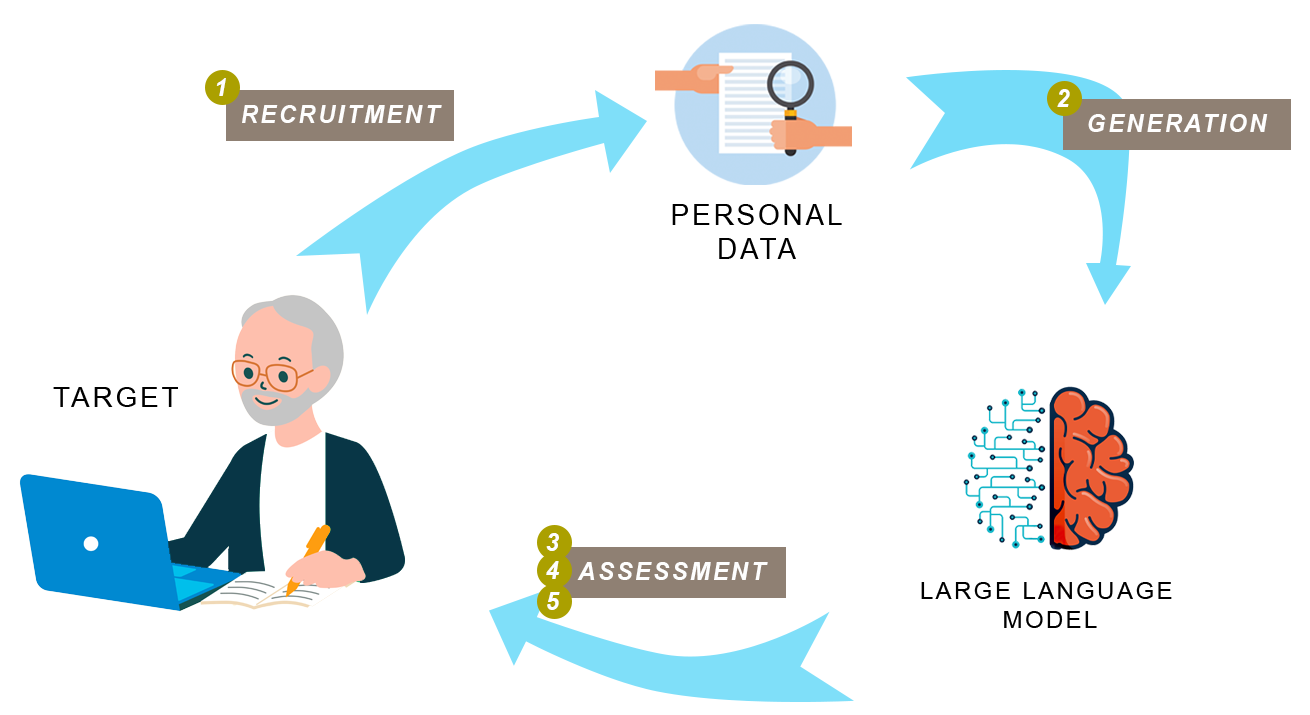}
  \caption{High-level workflow of the study: recruitment, collection of participant context, LLM-based message generation, and participant assessment.}
  \label{fig:study-workflow}
\end{figure*}

This section describes the project's rationale, design, and methods for data collection and analysis. We recruited working participants, collected information about their roles and workplaces, and used an LLM to generate spear phishing emails for participants to evaluate.  Our main outcomes were convincingness, stated response (open the link, investigate, delete, or report), and open-ended explanations. We treat the emails as simulated spear phishing messages, with the amount of personalized work-context detail included in pretexts varying by condition. We later added a post-hoc exploratory analysis of impersonation style, meaning how the LLM chose to portray the sender. Figure~\ref{fig:study-workflow} summarizes the study workflow.

\subsection{Design Rationale}
\label{sec:level}

Personalization level refers to the amount of recipient-specific workplace information included in a phishing pretext. We use it to measure personalized pretext depth. The study used four levels, from minimal tailoring to detailed workplace context:

\begin{itemize}
\setlength{\itemsep}{0pt}
\setlength{\parskip}{0pt}
\setlength{\parsep}{0pt}
\item \textbf{Basic (1)} – workplace referenced; generic greeting
\item \textbf{Low (2)} – basic level + participant's name and job title
\item \textbf{Moderate (3)} – low level + job responsibilities
\item \textbf{High (4)} – moderate level + coworker's name and shared-project details
\end{itemize}

Table~\ref{tab:synthetic-levels} uses fictional names to show which fields are added at each level. The scenario and sender remain fixed so the differences are easy to see. Participants did not see these examples. In the study, GPT-4o generated each message separately; the available profile fields were controlled by level, but scenario, tone, persuasion strategy, and sender configuration could vary across outputs. Four altered and deidentified historical outputs from one participant appear in Appendix~\ref{sec:app_historical_examples}; eight additional historical examples, two per level, are included in the reproducibility artifact.

\begin{table*}[t]
\centering
\caption{Researcher-constructed illustration of cumulative personalization depth. All entities are fictional; text shown was not presented to participants.}
\label{tab:synthetic-levels}
\small
\setlength{\tabcolsep}{4pt}
\begin{tabular}{p{0.08\textwidth}p{0.27\textwidth}p{0.58\textwidth}}
\toprule
Level & Information available & Illustrative addition to a fixed scheduling-workspace pretext \\
\midrule
1 & Organization & ``Northbridge Civic Services is moving the weekly scheduling workspace...'' \\
2 & Level 1 + recipient name and job title & ``Hello Alex, as an Operations Coordinator at Northbridge Civic Services...'' \\
3 & Level 2 + responsibilities & ``Because you coordinate weekly staffing schedules...'' \\
4 & Level 3 + coworker and project & ``Taylor mentioned that the Atlas records-migration pilot may affect the weekly staffing schedules...'' \\
\bottomrule
\end{tabular}
\end{table*}

The generation prompt in Appendix~\ref{sec:app_prompt} followed this cumulative field structure. These cumulative levels represent the amount of workplace information available to the model, not increasing levels of persuasion. We did not assign authority, urgency, social proof, reward, or politeness to particular levels. Because GPT-4o generated each message independently, these strategies and features could appear and vary at any level. We also left tone and sender role open so the model could construct the rest of the pretext. This choice allowed the LLM to select the sender persona (i.e., impersonation style), which we later analyzed in Subsection~\ref{sec:impersonation_style}.

A field experiment involving names, coworkers, and shared projects would require sending highly personalized deceptive messages to real workplaces. We instead used a disclosed survey so that participants could evaluate these messages without putting themselves, their coworkers, or their organizations at risk. This approach allowed us to compare responses across personalization levels, although it cannot show how people would behave in a real inbox. The study was designed to compare levels of personalization within simulated phishing pretexts, not to measure whether participants could distinguish phishing from legitimate workplace communication. We discuss this tradeoff along with other limitations further in Section~\ref{sec:limitations}.

\subsection{Participants}

Before recruitment, a conventional four-group analysis of variance (ANOVA) calculation ($\alpha=.05$, $1-\beta=.80$, $f=0.25$) suggested a target of approximately $N=180$ participants \cite{cohen1969}. We used this calculation only as a recruitment-planning benchmark; it was not a power analysis tailored to the repeated-measures mixed models used in the final analysis.

Participants were a self-selected convenience sample of working adults recruited through Prolific. Eligibility required current employment outside the home, U.S. residence, and no prior participation in a similar spear phishing message evaluation study. Eligible participants were directed to an online Qualtrics survey.

Of the 257 people who started the study, 67 timed out or did not finish. Of the remaining 190, six failed the attention check and were terminated without completing the survey. We excluded two responses completed in under 10 minutes and two with unusable workplace fields, leaving a final analytic sample of $N=180$. Copy/paste was disabled to deter scripted answers. Although the survey was designed to take approximately 20 minutes, the observed median completion time was approximately 30 minutes and 33 seconds.

Participants had a mean age of 37.0 years ($SD=11.7$, range 19--69). Eighty-nine (49.4\%) identified as female and 91 (50.6\%) as male. In total, 127 (70.6\%) reported full-time employment and 53 (29.4\%) part-time employment. Prior phishing-message exposure ranged from never (2.2\%) to daily (8.3\%), with the largest groups reporting exposure several times a week or about once a month (28.9\% each). Most participants were very (48.3\%) or somewhat (45.0\%) confident in detecting phishing.

\subsection{Ethics and consent}
All procedures received IRB approval (see Appendix~\ref{sec:app_ethics} for details). The Qualtrics survey began with a standalone consent page that participants had to accept to participate. The consent page described the study as voluntary, stated that participants could withdraw without penalty, instructed participants not to provide confidential or sensitive information, and disclosed that simulated spear phishing messages would be fictional and AI-generated. Only consenting U.S. working adults (18+) could proceed. Participant-provided identifiers and workplace-context fields were stored in restricted Qualtrics and university Box environments accessible only to the research team. The appendix documents data handling, retention, deidentification, and third-party processing in detail.

\subsection{Survey Procedure}
\label{sec:procedure}
After consenting, each participant completed a brief questionnaire covering prior exposure to spear phishing, self-reported confidence in identifying phishing messages, and demographics (age and gender). Participants then provided several fields: full name, workplace, job title, responsibilities, the first name of a known coworker, and a short description of a shared, non-confidential project. These fields were inserted into specific prompts via client-side Qualtrics JavaScript and transmitted to GPT-4o through four API requests. Each request asked for two messages at one personalization level. The survey was designed to produce eight messages per participant.

During data collection, four requested outputs (two Level~3 and two Level~4) were not returned. The generation code did not retry failed requests, so we excluded the corresponding four evaluations, leaving 1,436 valid message assessments across 180 participants. The resulting sample counts for each level are 360, 360, 358, and 358.

The messages were presented in random order within the survey. For each simulated email message, participants rated its convincingness on a 0--100 visual-analog scale. Participants were then required to select one stated action: open the link, investigate further, delete, or report as phishing. The item did not include an ignore option or allow multiple selections. Participants also rated their confidence in the selected action on a 0--100 scale. These were hypothetical survey responses; no links were active and no messages were delivered to real inboxes. After evaluating all messages, participants wrote open-ended explanations for the messages they found most and least convincing. These responses form the qualitative dataset.

\subsection{Impersonation style}
\label{sec:impersonation_style}
We use \textit{impersonation style} as shorthand for a nominal sender/pretext configuration. The code combines the persona of the apparent sender with whether the message refers to a known coworker. Two researchers coded each generated message after data collection. We define each style as follows:

\begin{itemize}
\setlength{\itemsep}{0pt}
\setlength{\parskip}{0pt}
\setlength{\parsep}{0pt}
\item \textbf{Style 1} – a department or entity is the apparent sender
\item \textbf{Style 2} – a named person not supplied as the coworker is the apparent sender
\item \textbf{Style 3} – a named person not supplied as the coworker references the supplied coworker in the message body
\item \textbf{Style 4} – the supplied coworker is the apparent sender
\end{itemize}

Impersonation style does not show whether the sender would actually be familiar, authoritative, or legitimate in the participant's workplace. To decide whether a message used the participant-provided coworker, coders traced the name to the supplied coworker field. This factor was not part of the initial design focus or independently randomized. The historical messages in Appendix~\ref{sec:app_historical_examples} provide observed examples. In the dataset, one Level~2 message happened to use the same common first name as a supplied coworker. We coded it as Style~2 because the coworker field was not available to the model at Level~2. No valid study message was coded as Style~4.

\subsection{Data Handling and Analysis}

We analyzed convincingness with a linear mixed-effects model and click intention with a frequentist logistic mixed-effects model fitted by maximum likelihood. Both models included a participant random intercept to account for repeated evaluations. The primary model specifications were:

\begin{align*}
\text{convincingness}
  &\sim C(\text{level}) + (1 \mid \text{participant}),\\
\operatorname{logit}\!\left[\Pr(\text{open-link})\right]
  &\sim C(\text{level}) + (1 \mid \text{participant}).
\end{align*}

Level~1 was the reference condition. The three contrasts with Level~1 use unadjusted confidence intervals and $p$-values; the separately reported 16 exploratory demographic tests use Holm adjustment. The primary models treated personalization level as a categorical within-participant factor; an ordinal linear-trend specification served as a sensitivity analysis. We summarized the post-hoc sender/pretext styles descriptively. We separately used an auxiliary multinomial logistic model with participant-clustered standard errors to examine investigate, report, and delete selections among responses that did not select open-link, using delete as the reference response. Demographic variables were examined as exploratory covariates and moderators.

The following definitions are used throughout the paper. 

\begin{description}
    \item[\textbf{convincingness}] Self-reported rating of how convincing each message seemed immediately after exposure, used as a survey measure of perceived credibility (integer $0$--$100$; higher $=$ more
  convincing).
    \item[\textbf{action}] Self-reported stated action: $open\ link$ (click), investigate further, delete, or report as phishing.
    \item[\textbf{click intention}] Selection of the open-link option in the survey. This is a stated response, not an observed click.
    \item[\textbf{personalization level}] Ordinal factor $\{1,2,3,4\}$ defined in Section~\ref{sec:level}, representing how much workplace detail was used in the pretext.
    \item[\textbf{impersonation style}] The post-hoc category $\{1,2,3,4\}$ defined in Subsection~\ref{sec:impersonation_style}, representing the sender/pretext configuration generated by the model.
\end{description}

\subsection{Qualitative Data Analysis}

We used inductive thematic analysis to study participants' free-text responses \cite{braun_using_2006}. Each participant provided four responses:

\begin{description}
\setlength{\itemsep}{0pt}
\setlength{\parskip}{0pt}
\setlength{\parsep}{0pt}
\item [\textbf{(i)}] why the top-ranked (most convincing) message was convincing,
\item [\textbf{(ii)}] why it was not convincing,
\item [\textbf{(iii)}] why the bottom-ranked (least convincing) message was convincing, and
\item [\textbf{(iv)}] why it was not convincing.
\end{description}

With four responses per participant, this yielded 720 open-ended explanations. All responses received at least one code. We used the non-thematic \texttt{Q0} label to responses that did not provide an interpretable explanation, which included unrelated feedback or other non-substantive text. Nineteen responses received only the \texttt{Q0} label and were excluded from thematic interpretation, leaving 701 responses in the thematic summaries.

Four researchers participated in the qualitative analysis workflow, with three of them serving as primary coders. The team first used open coding on a simple random sample of 103 responses (14.3\% of the corpus) to draft initial codes and rules. Coders marked every theme present in a response and assigned positive (+) or negative (-) valence. Codes were not mutually exclusive, so a response could receive several codes. Pairwise Cohen's $\kappa$ was calculated for each parent code on this independently coded subset, ignoring subcodes and valence. 

Disagreements involved whether a code applied, where related codes differed, which subcode fit, and whether several codes applied. The coders resolved these issues through discussion, codebook revision, and recoding, with input from the principal investigator when needed. This was done iteratively until the team reached a mean $\kappa$ of at least $.70$. The unweighted mean across seven estimable parent codes in the final iteration was $\kappa=.753$ (per-code range $.678$--$.951$). Neither coder assigned Communication Medium in this subset, so its chance-corrected $\kappa$ could not be calculated.One of the primary coders then coded the rest of the responses, with oversight from the rest of the research team. The final codebook contains eight main themes, with definitions and examples described in Section~\ref{sec:results}.

\section{Quantitative Results}
\label{sec:results}

We present the quantitative analyses across the 1,436 valid evaluations of AI-generated spear phishing messages. Participants rated convincingness at a mean of $60.2$ ($SD=29.6$) on the 0--100 scale and selected the open-link option in 36.9\% of responses. We report the percentages descriptively and use the models to compare outcomes across the tested pretexts.

\subsection{Effect of Personalization Level}

Table~\ref{tab:level} summarizes results by personalization level. Compared with Level~1, convincingness was 2.90 points higher at Level~2 (95\% confidence interval (CI) [$-0.60$, $6.39$], $p=.104$), 4.09 points higher at Level~3 (95\% CI [$0.59$, $7.59$], $p=.022$), and 7.61 points higher at Level~4 (95\% CI [$4.11$, $11.11$], $p<.001$) in the linear mixed-effects model. A sensitivity analysis treated the levels as a linear trend and estimated a 2.40-point increase for each additional level (95\% CI [$1.29$, $3.51$], $p<.001$).

\begin{table}[tb]
  \caption{Descriptive statistics by personalization level}
  \label{tab:level}
  \setlength{\tabcolsep}{4pt}
  \begin{tabular*}{\columnwidth}{@{\extracolsep{\fill}}lccc}
    \toprule
    Level & \shortstack{Convincingness\\$M\!\pm\!SD$} & \shortstack{Click\\intention (\%)} & Del:Rep \\
    \midrule
    Basic (1)    & 56.6 $\pm$ 31.8 & 31.9 & 0.37 \\
    Low (2)      & 59.5 $\pm$ 28.7 & 36.4 & 0.43 \\
    Moderate (3) & 60.7 $\pm$ 29.9 & 36.3 & 0.69 \\
    High (4)     & 64.3 $\pm$ 27.3 & 43.0 & 0.72 \\
    \bottomrule
  \end{tabular*}
  \footnotesize\par\medskip
  \textit{Note.} Level sample sizes are 360, 360, 358, and 358. Del:Rep is the delete-to-report ratio among responses that did not select open-link.
\end{table}

\begin{figure}[htbp]
    \centering
    \includegraphics[width=\linewidth,keepaspectratio]{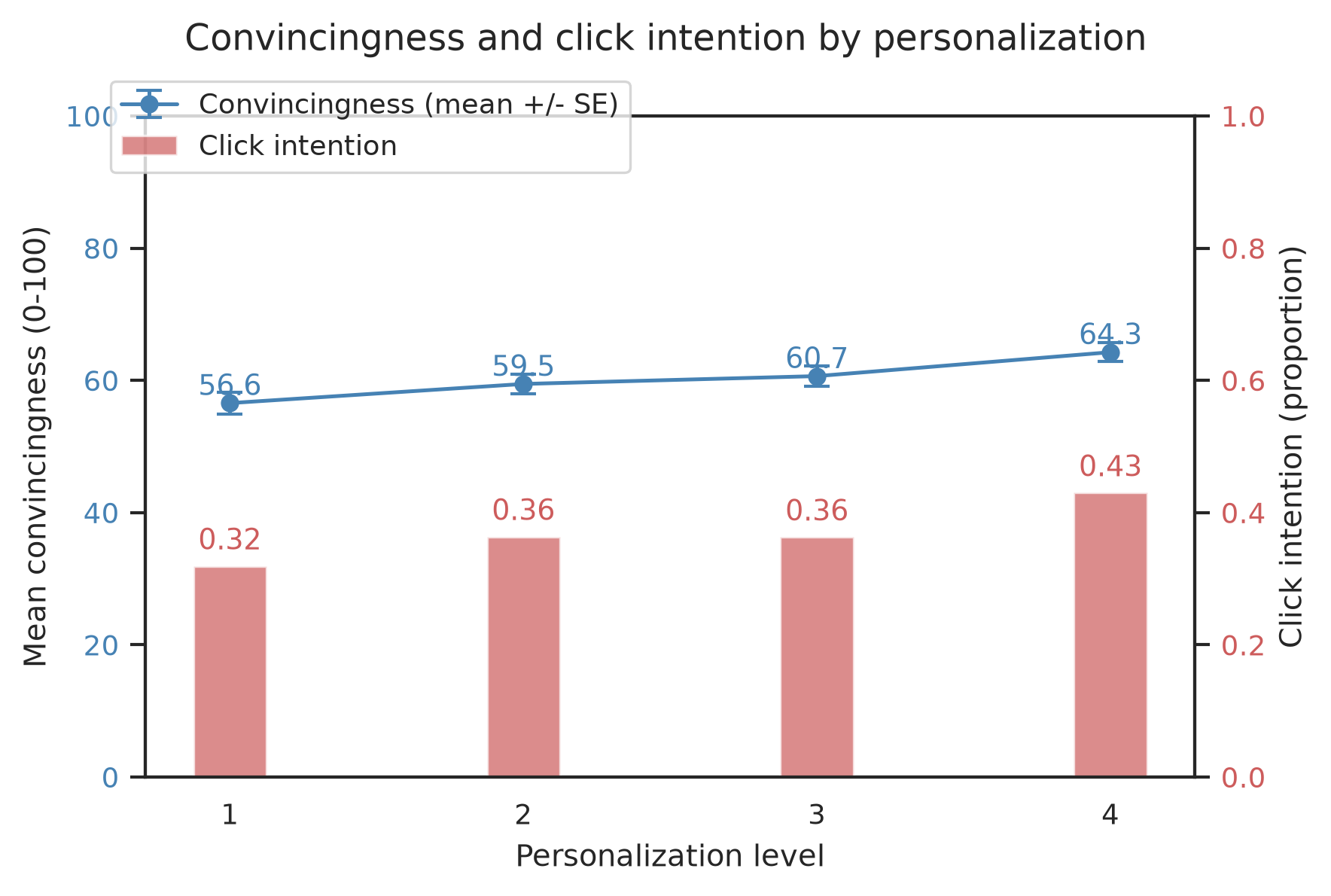}
    \caption{Convincingness and click intention by personalization level.}
    \label{fig_plevel}
\end{figure}

Click intention followed a similar pattern, with the clearest difference at Level~4. Compared with Level~1, the odds of selecting open-link were not clearly different at Level~2 (odds ratio (OR) $=1.41$, 95\% CI [$0.94$, $2.11$], $p=.099$) or Level~3 (OR $=1.39$, 95\% CI [$0.92$, $2.08$], $p=.114$), but were higher at Level~4 (OR $=2.29$, 95\% CI [$1.52$, $3.43$], $p<.001$). The ordinal sensitivity model estimated 28\% higher odds per level (OR $=1.28$, 95\% CI [$1.13$, $1.46$], $p<.001$). 

\begin{figure}[htbp]
    \centering
    \includegraphics[width=\linewidth,keepaspectratio]{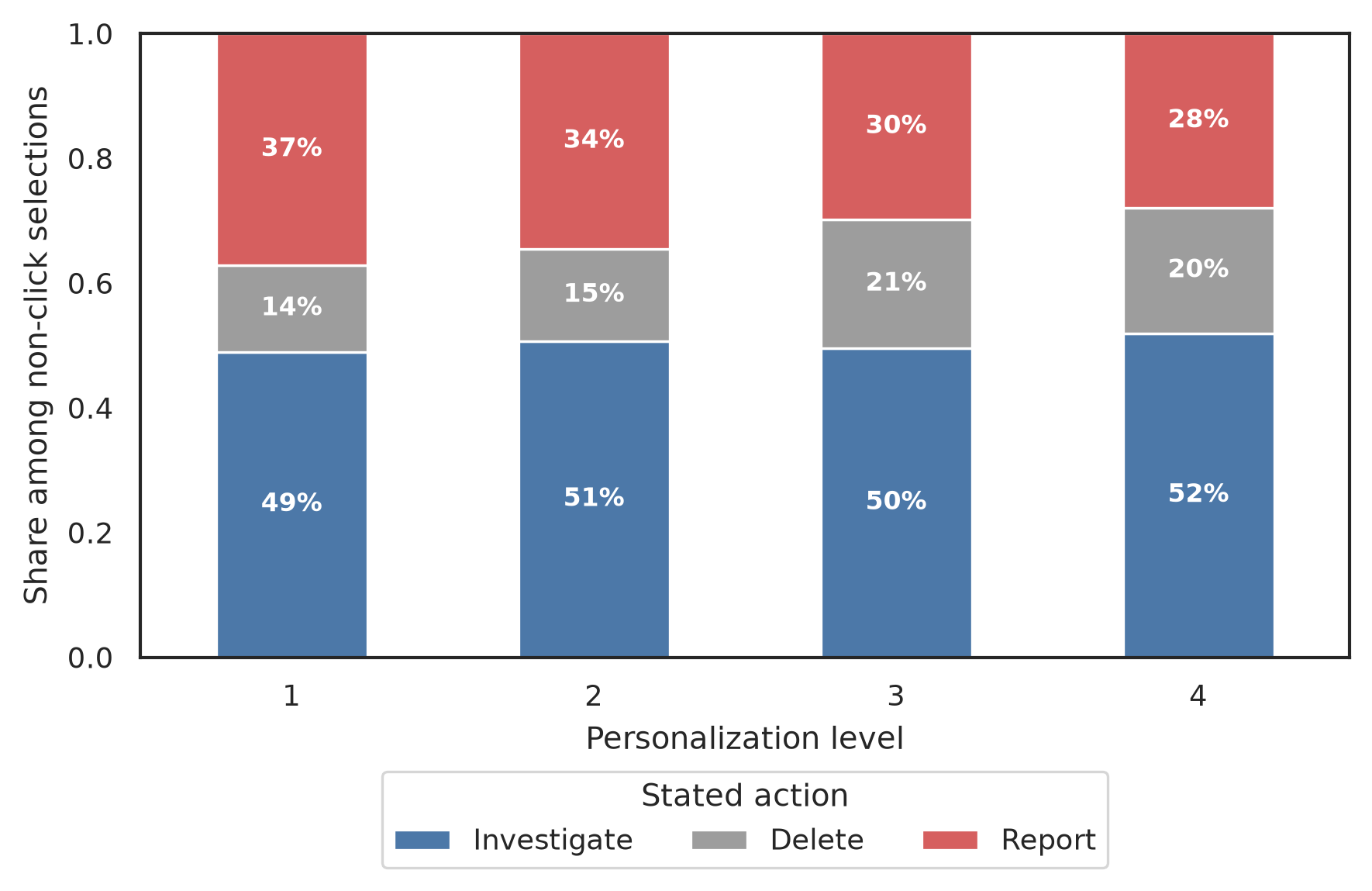}
    \caption{Distribution of investigate, delete, and report selections among responses that did not select open-link, by personalization level. Percentages are conditioned on non-click selections and may differ slightly from 100\% because of rounding.}
    \label{fig_pers_act}
\end{figure}

Figure~\ref{fig_pers_act} shows the other three choices among responses that did not select open-link. Investigation remained near half of selections (49--52\%). Reporting declined from 37\% at Level~1 to 28\% at Level~4, while deletion increased from 14\% to 20\%. An auxiliary clustered multinomial model compared these choices. For every one-level increase in personalization, the relative risk of selecting report rather than delete decreased by 21.7\% (relative risk ratio (RRR) $=0.78$, 95\% CI [$0.67$, $0.92$], $p=.002$). The estimate comparing investigation with deletion was less precise (RRR $=0.88$, 95\% CI [$0.77$, $1.01$], $p=.068$).

\subsection{Impersonation Style}

As an exploratory analysis, we examined the sender/pretext styles that GPT-4o produced, described in Table~\ref{tab:style}. Department/entity messages had the lowest convincingness and click intention (55.2; 31.0\%; $n=258$). Named-person messages came next (59.9; 35.1\%; $n=878$). Messages from a named person who referred to the participant-provided coworker had the highest values (65.7; 47.3\%; $n=300$). No valid message was coded as Style~4.

\begin{table}[tb]
  \caption{Descriptive statistics by impersonation style}
  \label{tab:style}
  \setlength{\tabcolsep}{4pt}
  \begin{tabular*}{\columnwidth}{@{\extracolsep{\fill}}lcc}
    \toprule
    Style & \shortstack{Convincingness\\$M\!\pm\!SD$} & \shortstack{Click\\intention (\%)} \\
    \midrule
    Department/entity (1) & 55.2 $\pm$ 32.9 & 31.0 \\
    Named person (2) & 59.9 $\pm$ 29.2 & 35.1 \\
    Coworker reference (3) & 65.7 $\pm$ 26.8 & 47.3 \\
    \bottomrule
  \end{tabular*}
  \footnotesize\par\medskip
  \textit{Note.} Style sample sizes are 258, 878, and 300. Style 4 was unobserved.
\end{table}

\begin{figure}[htbp]
    \centering
    \includegraphics[width=\linewidth,keepaspectratio]{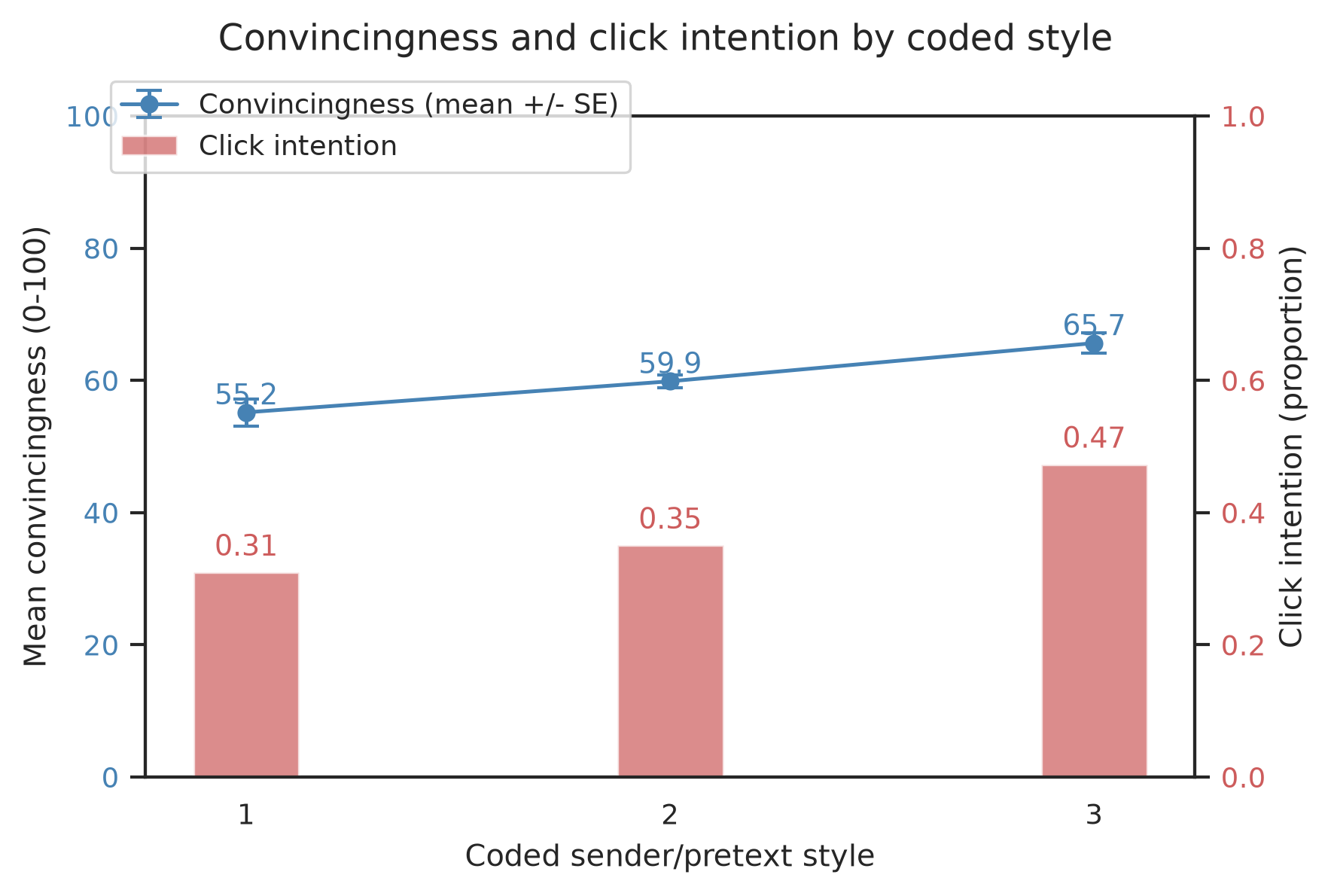}
    \caption{Convincingness and click intention by coded sender/pretext style. These descriptive differences do not isolate an effect of style.}
    \label{fig_style_conv}
\end{figure}

\begin{figure}[htbp]
    \centering
    \includegraphics[width=\linewidth,keepaspectratio]{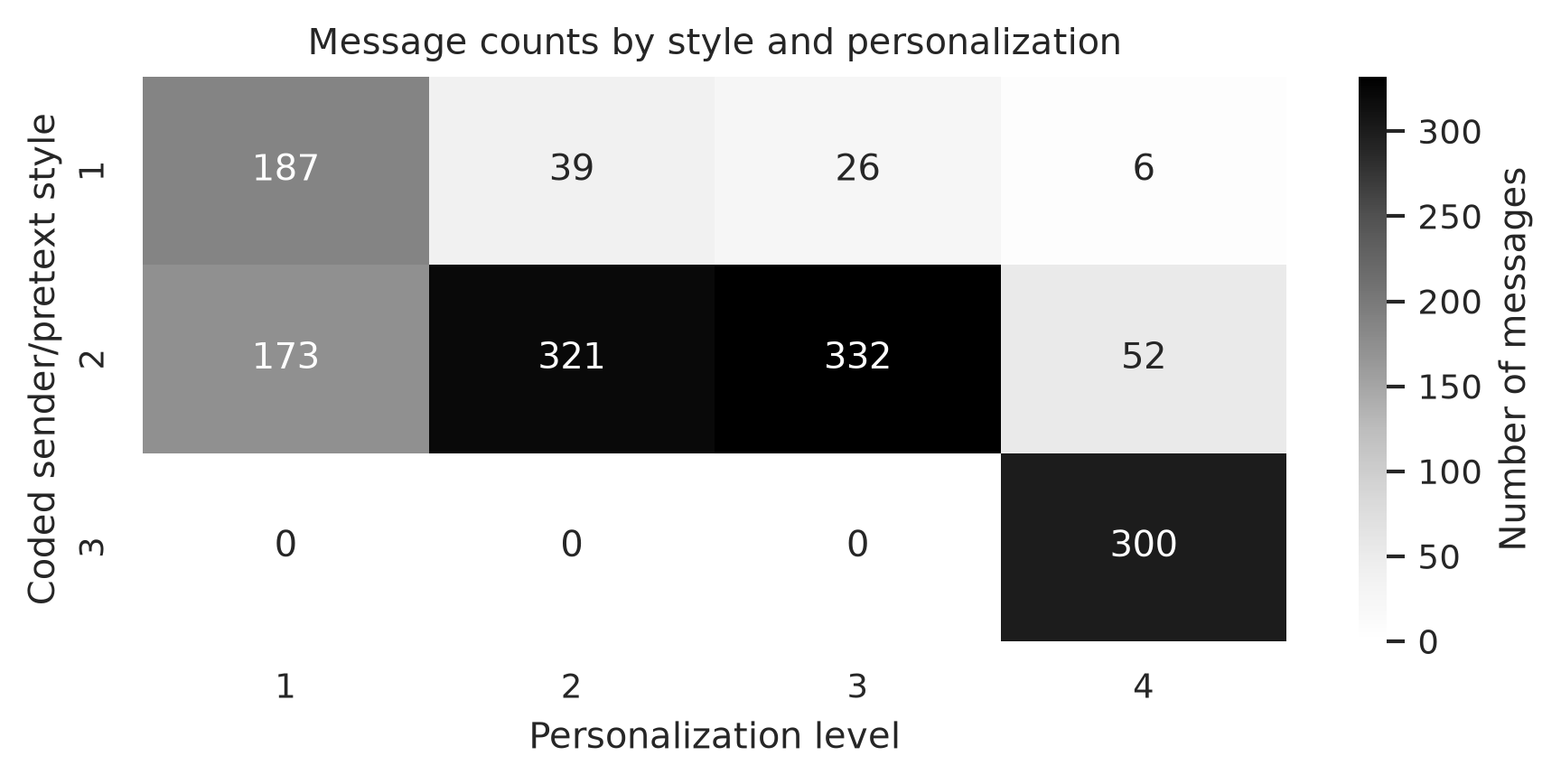}
    \caption{Message counts across personalization levels and coded sender/pretext styles. Empty Style~3 cells make the full categorical interaction non-estimable.}
    \label{fig_heat_count}
\end{figure}

This style analysis was post-hoc and not independently randomized, and styles were strongly imbalanced across personalization levels. All 300 Style~3 messages occurred at Level~4; no Style~3 messages occurred at Levels~1--3. We therefore could not estimate a full categorical Style $\times$ Personalization interaction. Figure~\ref{fig_heat_count} shows this imbalance. Because only Level~4 supplied the coworker field, any person names at lower levels came from the model itself. We therefore present the style differences only descriptively and do not treat them as independent effects of sender style.

\subsection{Other Factors}

We explored whether age, gender, prior phishing-message exposure, or confidence in detecting phishing predicted the outcomes or changed the relationship with personalization. None of the 16 overall tests for convincingness or click intention remained significant after Holm correction (all adjusted $p\geq.795$). This does not prove that the subgroups responded identically.

\section{Qualitative Results}
\label{sec:qual_results}
This section explains why participants found messages convincing or suspicious. Our thematic coding produced a codebook with eight main themes. As mentioned in Section~\ref{sec:method}, nineteen responses received only the non-thematic \texttt{Q0} label because they were non-substantive or uninterpretable, and thus were excluded.

For each theme below, we provide its working definition, common indicators, and some illustrative quotes. Participant identifiers (e.g., R1) indicate which participant said each quote. Names of real people and workplaces are redacted. We then report how often each theme appeared and whether comments about it were more often positive or negative. Table~\ref{tab:qual_codes} gives a short description of each code.

\begin{table}[tb]
  \caption{Qualitative codes and descriptions}
  \label{tab:qual_codes}
  \setlength{\tabcolsep}{3pt}
  \small
  \begin{tabular}{>{\raggedright\arraybackslash}p{0.30\columnwidth}>{\raggedright\arraybackslash}p{0.63\columnwidth}}
    \toprule
    \textbf{Name} & \textbf{Description}  \\
    \midrule
    Personal & Message feels relevant/mismatched to the participant's current work context \\
    Style & Message tone, clarity, formality, and formatting \\
    Source & The sender identity and authority cues  \\
    Requested Action & Link/file interactions, routine requested actions, or sensitive/high-risk actions \\
    Urgency & Pressure/threats to act quickly  \\
    \shortstack[l]{Benefit/Favor/\\Reward} & Benefits, opportunities, favors, incentives, gifts, or ego appeals \\
    Medium & The message channel matches/mismatches expectations  \\
    General & Unsupported overall credibility judgments when no substantive theme applies \\
    \bottomrule
  \end{tabular}
      \footnotesize\par\medskip
\textit{Note.} \textit{Personalization} is abbreviated as \textit{Personal}.
\end{table}

\par\smallskip\noindent\textbf{Personalization / Relevance.}
Participants linked credibility of a message to how closely it fit their daily work. Among 329 valenced Personalization/Relevance mentions, 220 (66.9\%) were positive and 109 (33.1\%) were negative. This total is four higher than the 325 responses containing the theme because four responses included both positive and negative personalization subcodes. Messages that combined personal details---full name, job title, employer, coworker, or current project---were often mentioned as convincing. Messages were described as convincing because they referenced “something I would be doing often in my job” (R11) or “my company’s name and also \ldots my job role \ldots” (R164). References to specific work (e.g., “licensing system project” (R164), “informative session” (R16), “ongoing campaign strategies” (R36)) and routine tasks like scheduling---“one of the most integral parts of my job and that part got to me most” (R134)---were also deemed more convincing.

Negative mentions described pretexts that did not fit the participant's real work. Often this occurred because details were simply wrong: one message “said for me to grab my lesson plans, [but] I don’t do lesson plans” (R15), and another participant said, “[I] don’t have an events coordination team” (R59). Other messages were less convincing because the personalized context was too vague: one message “lack[ed] specific details about the\ldots purpose or context” (R115), and another noted a “lack of specific detail about the conference” (R61). Overall, accurate details that aligned with participants' work were more persuasive, while incorrect or vague details created mismatches.

\par\smallskip\noindent\textbf{Message Style.}
Tone and styling shaped credibility in several ways. Among valenced Message Style mentions, 46.6\% were positive and 53.4\% were negative. Convincing messages used a “casual tone” (R87) and felt “naturally friendly\ldots not over the top” (R83). Messages seemed genuine when their style matched local norms in their workplace. One participant noted that “it was casual, which is typical of our office” (R31), while others said it “looked professional and relevant to my role,” “used a familiar tone” (R53), or resembled “typical corporate communication” (R76).

Negative Message Style mentions described messages as “overly flattering” (R88), “a bit too friendly” (R73), or sounding “cliché,” “boilerplate” (R98), or “mechanical” (R91). Stock phrases like “I hope this finds you well” or “best regards” made one message “more than likely a scam” (R34). Greetings that were too formal or too informal, such as “Cheers” (R13), or “way too white collar for the people I am interacting with” (R14) were also deemed suspicious. Overall, when style cues matched the tone of the workplace, either casual or professional, the messages were more persuasive; when they were mismatched or looked like stock phrases, they were flagged as suspicious.

\par\smallskip\noindent\textbf{Source.}
The source of a message, or the perceived authority of its sender, also mattered. Among valenced Source mentions, 42.1\% were positive and 57.9\% were negative. One participant noted that “the sender is from the [company name] security team. This gives the air of authority and that I should listen to her and follow her instructions” (R20). Those who were convinced explained it was in part due to the titles and roles used in messages such as “a safety compliance officer” (R22), “Investment Analyst” (R64), and senders from clients or internal departments (“human resource department” (R85); “a trusted IT team member connected to my department” (R103)).

Messages were less convincing when they arrived “without providing a means of confirming its legitimacy, such as mentioning an internal portal or HR contact, and the sender’s name” (R114), or from an “unfamiliar sender” or “vague source” (R157) with “no prior communication” (R148). Overall, clear and plausible authority cues raised persuasiveness, while the lack thereof raised suspicion.

\par\smallskip\noindent\textbf{Requested Action.}
Participants also judged what the message asked them to do, including opening links or files, completing routine tasks, and taking sensitive or risky actions. Among valenced Requested Action mentions, 23.6\% were positive and 76.4\% were negative. One message felt like “a serious announcement made for a thorough reason” (R90); a breach notice was “very harmful to [their] career and the company’s well being,” making it “very convincing to prompt me \ldots to secure my account” (R46). Thus, a plausible context could make some requested actions appear credible.

More often, participants treated the requested action as an indicator of a fake message. Requests for passwords, sensitive documents, or credential checks were considered red flags: “any message to verify credentials is probably \ldots a phishing attack” (R106); “we are not sent emails to update the accounts ourselves” (R123). Routine requests also reduced credibility when they did not fit the participant's role or normal workflow. Overall, requested actions were often suspect, but a compelling context could make them believable.

\par\smallskip\noindent\textbf{Urgency / Pressure.}
Urgency and pressure cut both ways. Among valenced Urgency/Pressure mentions, 52.4\% were positive and 47.6\% were negative. For some, a deadline added realism. One participant said, “failure to do so in 48 hours will lead to delays” “got me thinking” (R146). Another participant highlighted an email from “the ‘IT team at [company name]’” because of its “very professional appearance” and “sense of urgency” (R138). Others saw pressure as a warning: “Once i spot pressure, i become much aware of the possibilities of something fishy. Why must this be urgent? why the pressure?” (R54). Another wrote that “the urgency with which it encourages action to be taken is quite off for me” (R110). Overall, urgency could signal importance or harmful intent, depending on the reader.

\par\smallskip\noindent\textbf{Benefit / Favor / Reward.}
Participants also noticed benefits, opportunities, favors, rewards, and appeals to ego. Among valenced Benefit/Favor/Reward mentions, 75.8\% were positive and 24.2\% were negative. These cues could make a message seem legitimate and motivate action: “The incentive of a \$100 gift card added a sense of legitimacy and motivation” (R53), and “[the] request for photos also sounded like a genuine favor from a colleague” (R87). They could also raise suspicion because “unsolicited offers of gifts are common in phishing and spam” (R148), or because a “reward element felt a bit too promotional for a real internal survey, which raised mild suspicion” (R53). Overall, participants more often described these cues as increasing credibility or motivation than as raising suspicion.

\par\smallskip\noindent\textbf{Communication Medium.}
Communication channel was mainly a negative cue. Among 62 valenced Communication Medium mentions, 58 (93.5\%) were negative and four (6.5\%) were positive. One participant would “normally go look in [Microsoft] Teams rather than click on a link to the document” (R13). Another said, “I would likely receive a call from [company name] if this was the case” (R20). Personal invites were expected via call or text (“catch up with me for coffee” (R22)); networking events were “not via email” (R96). Overall, channel mismatches generally made messages less convincing and more suspicious.

\par\smallskip\noindent\textbf{General.}
A small group of responses gave an overall judgment without a specific reason, such as “Honestly, nothing at all” (R83), “it doesn’t convince me” (R73), and “There was no aspect that made it less convincing” (R134). We placed these responses in the residual General category. Among 22 valenced General mentions, 12 (54.5\%) were positive and 10 (45.5\%) were negative.

\subsection{Descriptive Coverage of Qualitative Codes}
\label{sec:qual-descriptives}

This subsection summarizes how often each category appears in the open-ended responses and how those codes are distributed across \textit{Convincing} vs.\ \textit{Less convincing} explanations and \textit{Top} vs.\ \textit{Bottom} ranked messages. Unless stated otherwise, each percentage is the share of distinct retained responses in which a parent code appeared at least once. Table~\ref{tab:coverage-merged} details the coverage of the codes across the dataset and within the defined strata.

\begin{table}[t]
\centering
\caption{Coverage of qualitative codes across $N{=}701$ retained responses: overall (\% of responses with code present) and within Convincing/Less-convincing and Top/Bottom strata.}
\label{tab:coverage-merged}
\resizebox{\linewidth}{!}{%
\begin{tabular}{lrrrrrrr}
\toprule
\textbf{Code} & \textbf{$n$ responses} & \textbf{\% overall} & \textbf{Conv \%} & \textbf{Less \%} & \textbf{Bottom \%} & \textbf{Top \%} \\
\midrule
Personal & 325 & 46.4 & 61.4 & 31.2 & 44.9 & 47.9 \\
Style & 325 & 46.4 & 45.5 & 47.3 & 47.7 & 45.0 \\
Action & 161 & 23.0 & 14.2 & 31.8 & 22.3 & 23.6 \\
Source & 145 & 20.7 & 16.8 & 24.6 & 19.1 & 22.2 \\
Benefit & 95 & 13.6 & 19.6 & 7.4 & 14.3 & 12.8 \\
Urgency & 82 & 11.7 & 11.6 & 11.7 & 9.7 & 13.7 \\
Medium & 62 & 8.8 & 2.8 & 14.9 & 8.3 & 9.4 \\
General & 22 & 3.1 & 2.3 & 4.0 & 3.7 & 2.6 \\
\bottomrule
\end{tabular}%
}
\footnotesize\par\medskip
\textit{Note.} All percentages are shares of retained responses. Convincing ($n=352$), Less-convincing ($n=349$), Bottom ($n=350$), and Top ($n=351$) columns report coverage within those strata.
\end{table}

\par\smallskip\noindent\textbf{Overall coverage.}
Among the 701 responses, Personalization/Relevance and Message Style appeared most often (325 responses each; 46.4\%). Requested Action appeared in 161 responses (23.0\%) and Source Cues in 145 (20.7\%). Each remaining theme appeared in 3.1--13.6\% of responses. Because responses could receive more than one code, these percentages do not sum to 100\%.

\par
\begin{figure}[htbp]
  \centering
  \includegraphics[width=0.95\linewidth]{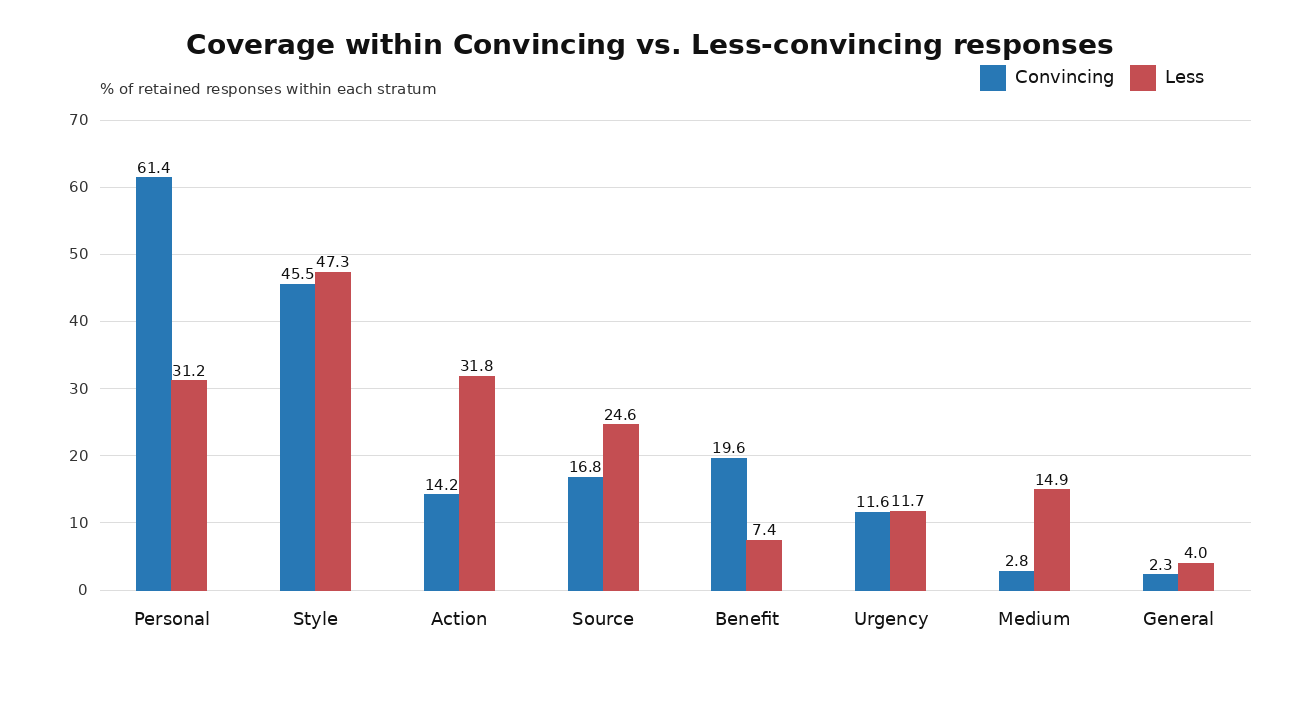}
  \caption{Coverage by Convincing vs.\ Less-convincing explanation (\% within each response stratum).}
  \label{fig:convless}
\end{figure}

\smallskip\noindent\textbf{Convincing vs.\ Less convincing.}
Explanations for \textit{Convincing} messages emphasized Personalization more than explanations for \textit{Less convincing} messages (61.4\% vs.\ 31.2\%; $\Delta{=}30.2$ percentage points) (Figure~\ref{fig:convless}). This pattern complements the quantitative finding by showing that personal details were discussed more often in explanations of convincing messages. However, the presence of Personalization in 31.2\% of \textit{Less convincing} explanations also shows that personalization was not uniformly helpful. Some participants treated mismatched or incomplete personal details as reasons for suspicion. Benefit/Favor/Reward also appeared more often in \textit{Convincing} explanations (19.6\% vs.\ 7.4\%). In contrast, Communication Medium and Requested Action were more common in \textit{Less convincing} explanations (Table~\ref{tab:coverage-merged}). Even a personalized message could lose credibility when its channel or requested action did not fit expectations.

\begin{table}[htbp]
\centering

\caption{Net persuasion by code (valenced mentions only).}
\label{tab:net}
\begin{tabular}{lrrrrr}
\toprule
\textbf{Code} & \textbf{+} & \textbf{-} & \textbf{Total} & \textbf{Net} & \textbf{\%Net} \\
\midrule
Benefit & 72 & 23 & 95 & 49 & 51.6 \\
Personal & 220 & 109 & 329 & 111 & 33.7 \\
General & 12 & 10 & 22 & 2 & 9.1 \\
Urgency & 43 & 39 & 82 & 4 & 4.9 \\
Style & 152 & 174 & 326 & -22 & -6.7 \\
Source & 61 & 84 & 145 & -23 & -15.9 \\
Action & 38 & 123 & 161 & -85 & -52.8 \\
Medium & 4 & 58 & 62 & -54 & -87.1 \\
\bottomrule
\end{tabular}
\end{table}

\par\smallskip\noindent\textbf{Net persuasion index.}
To summarize whether a theme was described as \emph{helping} or \emph{hurting} persuasion, we calculated a net persuasion index from the positive and negative mentions of each code. A response could contribute both a positive and a negative mention. A positive score means the theme more often made a message convincing, while a negative score means it more often raised suspicion. Formally,

\begin{equation*}
\%Net = 100 \cdot \frac{\#(+) - \#(-)}{\#(+) + \#(-)} .
\end{equation*}

\begin{figure}[htbp]
  \centering
  \includegraphics[width=0.95\linewidth]{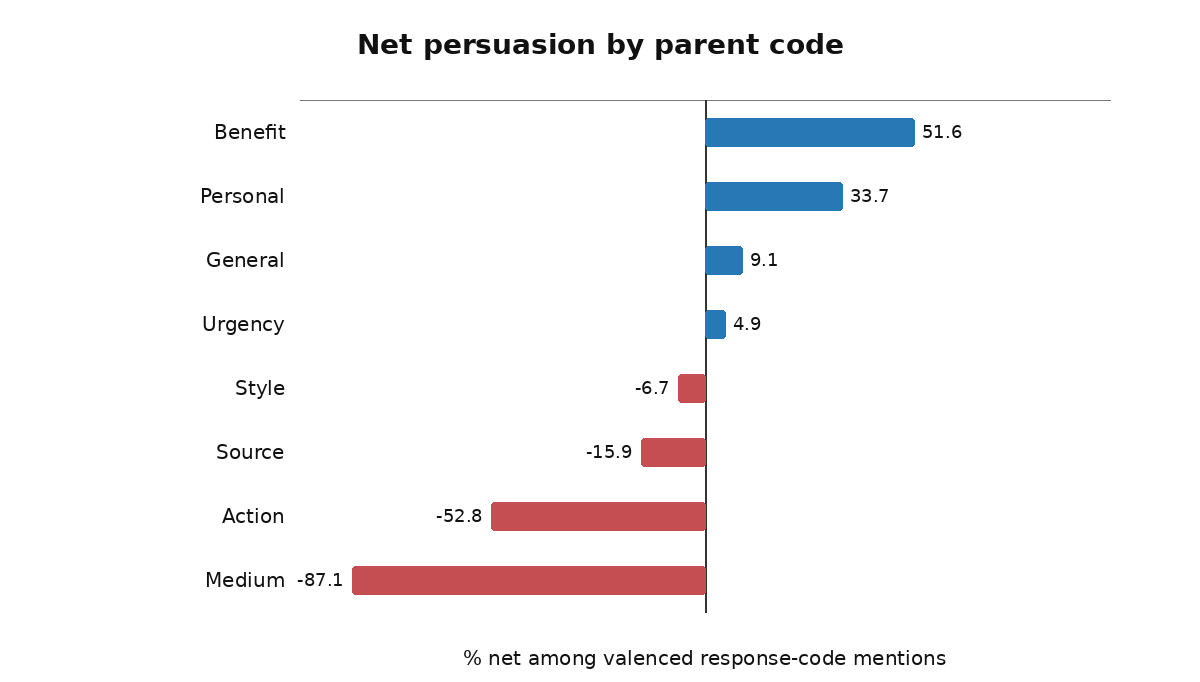}
  \caption{Net persuasion index by code.}
  \label{fig:netpersuasion}
\end{figure}

Table~\ref{tab:net} and Figure~\ref{fig:netpersuasion} detail the net persuasion index for each code. Benefit (+51.6\%) and Personalization (+33.7\%) skew positive when participants commented on them. The Personalization score should be read together with its negative mentions: personalization was usually helpful when it fit the recipient's work context, but it also raised suspicion when details were wrong, vague, or implausible. Urgency is near neutral (+4.9\%), while Message Style leans slightly negative ( $-6.7\%$). Requested Action ($-52.8\%$) is strongly negative, and Communication Medium ($-87.1\%$) was the strongest negative cue. General is residual-only and should not be interpreted as comparable to the other themes.

\section{Discussion}
\label{sec:discussion}

This section discusses how participants judged the simulated spear-phishing messages and what they intended to do. We consider how more personalized details can make a message seem more believable but can also raise suspicion when those details do not fit the workplace context. We also discuss coworker references, individual differences, and implications for verification and reporting. Throughout, we interpret the results as relative differences in survey responses across participants.

\par\smallskip\noindent\textbf{Personalized pretexts increased convincingness.}
Convincingness rose as personalization increased from Level~1 to Level~4. The percentage expressing click intention also rose descriptively, from 31.9\% to 43.0\%. The categorical models indicate that evidence was strongest at Level~4 rather than showing a clear difference at every intermediate level. This pattern agrees with laboratory and field findings that role, coworker, and project references can make phishing emails seem better matched to a recipient \cite{jagatic2007social,xu_personalized_2023,czybik_personalized_2026}. Among responses that did not select open-link, investigation remained common at every level. This suggests uncertainty rather than simple acceptance or rejection. Higher personalization was also associated with lower reporting relative to deletion in our model. Whether this pattern would reduce organizational awareness in a real deployment remains a question for future research.

\par\smallskip\noindent\textbf{Coworker references worked without direct impersonation.}
Our prompt allowed the LLM to choose the sender persona. Instead of presenting a supplied coworker as the sender, the model usually used that coworker as supporting context. A named person or organizational entity referred to the coworker to support the pretext. No valid message directly impersonated the coworker. One Level~2 message happened to use the same common first name, but the coworker field was not available to the model in that condition, so the message was therefore coded as Style~2. One possible explanation for this pattern is that model safety guardrails may have discouraged direct impersonation, but our data cannot establish that explanation. Another is that the prompt did not explicitly state the coworker to be the intended sender, so it did not consider the scenario. Messages from a named person who referenced a coworker (Style~3) had a higher percentage expressing click intention than messages from a department or entity (47.3\% vs.\ 31.0\%), but all Style~3 messages occurred at Level~4, so this descriptive relationship is mixed with personalization level and other message features.

These findings are consistent with evidence that authority and role cues influence clicking and reporting in workplace phishing \cite{Williams2018IJHCS,Steves2020PhishScale} and with guidance describing common BEC tactics that exploit organizational roles and familiar relationships \cite{FBI_BEC,CISA_TA15_BEC}. Impersonation style may therefore be a separate part of a pretext alongside personalization level. Future work should manipulate sender identity directly in a more balanced design. Direct impersonation may be more persuasive when accurate, but more suspicious when the impersonation includes incorrect details.

\par\smallskip\noindent\textbf{Individual differences were limited.}
We also examined whether age, gender, phishing exposure, or self-reported confidence changed how participants responded to the messages. We did not find reliable main or moderation effects after correcting for multiple comparisons (all adjusted $p\geq.795$). These null results do not prove that participant groups responded in the same way. The study may not have had enough power to detect smaller subgroup differences. Future work with larger and more diverse samples could better test whether personalization and impersonation affect particular populations or experience levels differently.

\par\smallskip\noindent\textbf{Context fit shaped persuasion and suspicion.}
The open-ended responses help explain why richer personalization was associated with higher convincingness and more click intention. The most frequent themes were Personalization/Relevance and Message Style (46.4\% each), with Personalization appearing more often in \emph{Convincing} explanations than in \emph{Less-convincing} ones (61.4\% vs.\ 31.2\%). Participants described messages as more convincing when details aligned with their role, responsibilities, or active projects, and became suspicious when mismatches existed. This reasoning is consistent with prior work showing that recipients rely on contextual signals under uncertainty and that attackers can exploit those signals to be more persuasive \cite{benenson_unpacking_2017,oliveira_empirical_2019,Dhamija2006WhyPhishingWorks}. Message Style (tone, clarity, and formatting) helped when it matched the participants' workplace norms, while boilerplate or overly flattering phrasing did the opposite. Participants also referenced Source cues, where job titles, departments, or implied authority affected their willingness to comply, aligning with findings that role and authority framing influence workplace phishing responses \cite{Williams2018IJHCS,Steves2020PhishScale}.

The qualitative data also show that personalized cues were not always persuasive. Among valenced Personalization/Relevance mentions, 33.1\% were negative, usually because the pretext failed to match the participant's real work context. Incorrect details, vague references, or details that were too hard to verify made the message easier to reject. Thus, personalization worked through fit , not simply through the number of details. Personalized details helped more often than they hurt, but mismatches remained an important failure mode. The net persuasion results reinforce this pattern. Benefit and Personalization skewed positive, Communication Medium and Requested Action skewed negative, and Message Style, Source Cues, and Urgency depended more heavily on context.

Requested actions (e.g., credential checks, document requests, or account verification) were frequently treated as suspicious even when the message looked professional. This agrees with classic findings that people use both surface cues and context to judge legitimacy \cite{Dhamija2006WhyPhishingWorks}. Communication Medium was similarly important. Participants noted that legitimate requests would normally arrive through a different channel, such as Teams, phone, or an internal portal, and that email links felt inappropriate for certain actions. This points to a limitation of LLM-generated pretexts: even when a model can insert plausible personal details, it may not reliably infer which channel or request format is normal in a particular workplace. A request that seems reasonable in Teams, by phone, or through an internal portal may become suspicious when it arrives as an email link.

\par\smallskip\noindent\textbf{Training should target verification and reporting.}
Our results point to the need to both reduce clicking \emph{and} increase reporting. Higher personalization increased convincingness and click intention, and among non-click responses it was associated more clearly with declining reporting than with declining investigation. Training should emphasize that personal details do not prove that a message is legitimate. It should also teach simple verification habits, such as using known portals or known contact methods \cite{canova_nophish_2015}. Recent work also shows that matching phishing training to a user's proficiency can improve training outcomes, although our study did not test a training intervention \cite{schoeni_training_2025}. Future work should test whether training examples matched to employees' roles and communication channels help them recognize both accurate pretexts and mismatches. Reporting should also be quick and easy so that employees report suspicious messages even when they do not open the link.

A practical next step is to compare role-tailored training with generic training. Survey studies can measure stated responses, whereas field studies can measure observed clicking and reporting. LLMs may help generate tailored training examples efficiently, but researchers still need to review them to avoid unsafe or overly invasive content.

\section{Limitations and Future Work}
\label{sec:limitations}

We enumerate the limitations of this study, as well as suggestions for future research. First, this study is best suited to comparing the four personalization conditions. Each participant rated messages from every level in the same disclosed survey. This design allowed us to compare responses as the model received more workplace context, but it cannot estimate real-world click-through rates. It also allowed participants to assess highly personalized messages without exposing them, their coworkers, or their organizations to deceptive messages in a real workplace.

Second, our measures rely on self-report, which limits how directly they represent real behavior. Convincingness captures perceived credibility, while the open-link response captures click intention. Prior phishing research has used laboratory and survey measures to study detection, suspicion, and stated responses under controlled conditions \cite{hakim_phishing_2021,xu_personalized_2023}. We follow this approach and treat our outcomes as comparative survey measures rather than real-world behavior. Participants might respond differently in a workplace, especially because they knew they were evaluating simulated spear phishing. The phishing-awareness questions presented before the message evaluations may also have encouraged closer scrutiny than an everyday inbox encounter. As explained in Section~\ref{sec:level}, this is a tradeoff of the disclosed survey design. These survey cues may have affected the overall responses, even though every personalization condition was evaluated using the same procedure.

Third, the survey setting has limited ecological validity. Messages appeared in Qualtrics rather than in participants' inboxes, and the survey did not reproduce real links, attachments, warning banners, email-client interfaces, or organizational reporting policies. It also did not include legitimate workplace messages, so the study cannot show how well participants distinguish phishing from normal communication. Creating legitimate control messages would require either authentic workplace communications, which could create additional privacy concerns, or separately validated synthetic messages. Generating them from the same profile information would not establish that they reflected normal communication in each participant's workplace. The forced-choice response also omitted an ignore option and did not allow sequential or combined actions. Together, these omissions limit how closely the study represents inbox behavior. Future work should test these findings in live or longitudinal settings that measure observed clicking.

Fourth, our participants were a self-selected convenience sample of U.S. working adults recruited through Prolific. They may not represent workers in other countries, cultures, or organizational settings. Future studies should test whether the findings generalize to these populations and settings.

Fifth, we studied email only. Future work should examine whether personalization operates differently in SMS, workplace messaging platforms, and attacks that span multiple communication channels.

Sixth, impersonation style was examined in a post-hoc exploratory analysis. Our coding describes how the generated message presented its sender and pretext, but it cannot establish whether that sender would actually seem familiar or legitimate in the participant's workplace. Impersonation styles were also not balanced across personalization levels. We therefore cannot separate the influence of impersonation style from personalization and other message features. Future work should use a fully crossed design that independently controls sender identity and personalization level.

Finally, the full qualitative corpus was not independently double-coded. The reported agreement applies only to the 103-response codebook-development subset. Future work should independently double-code the full corpus or a larger subset using the finalized codebook.

\section{Conclusion}
This study examined how participants responded to LLM-generated spear phishing pretexts containing different levels of workplace personalization. In our disclosed survey, the strongest differences appeared at Level~4: participants rated the most personalized messages as more convincing and had higher odds of expressing click intention than at Level~1. Personalization was not always persuasive. Details that fit a participant's role, routines, and workplace context made messages seem more credible, while incorrect, vague, or inappropriate details raised suspicion. Messages that referenced a coworker also received higher convincingness ratings and a higher click intention. However, all of these messages appeared at Level~4, so a balanced experiment is needed to separate the influence of coworker references from personalization level. Among responses without click intention, investigation remained common, while reporting became less likely relative to deletion as personalization increased. Although these results do not estimate real-world clicking or compromise, they show why personalized details should be treated as claims to verify rather than proof of legitimacy. Security training should reinforce this habit, and organizations should make suspicious messages easy to report.
\appendices

\section{Ethical Considerations}
\label{sec:app_ethics}

\paragraph{Oversight, consent, and research subject population.}
All procedures were conducted under institutional review board (IRB) approval (IRB Number: IRB2025-128). Research subjects provided implied consent via a standalone Qualtrics consent page. The consent page stated that participation was voluntary, subjects could withdraw at any time without penalty, and subjects should not reveal confidential or sensitive information. Recruitment targeted U.S.-based working adults (18+) (through Prolific), excluding minors and other vulnerable populations by design. Subjects recruited via Prolific were compensated \$5.

\paragraph{Consent materials.}
Because the full form contains operational contact details and vendor-policy language, we summarize the consent content here.The consent page disclosed that subjects would provide profile and workplace-context information, evaluate simulated spear phishing messages, and assess message convincingness. It also stated that the messages were fictional and AI-generated, that GPT-4o would be used through the vendor API policy, and that the study involved privacy risk from collection of digital records. The consent materials described secure storage in Qualtrics and a restricted university Box environment and three-year data retention.

\paragraph{Stakeholders.}
Primary stakeholders include:
\begin{description} 
    \item[\textbf{(i)}] subjects, whose safety and privacy we must protect,
    \item[\textbf{(ii) }] organizations and third parties referenced in subject-provided data (e.g. coworkers), who could face risk if identifiable information were disclosed, and
    \item[\textbf{(iii)}] service providers used for recruitment, survey delivery, and LLM message generation.
\end{description}
A secondary stakeholder is potential attackers, insofar as our findings about persuasive cues could be misused.

\paragraph{Data minimization and third-party references.}
To generate simulated spear phishing messages, subjects provided limited work-context fields: name, workplace/employer, job title, job responsibilities, the first name of a coworker, and a short description of a shared non-confidential project. We explicitly instructed subjects not to enter sensitive information and requested that any coworker reference be first name only, along with a non-confidential project context. We treated coworker names as third-party personal information requiring additional protection.

\paragraph{Risks and mitigations.}
The study was classified as minimal risk. Foreseeable risks included 
\begin{description} 
    \item[\textbf{(1)}] privacy risk from collection of identifiers and work context,
    \item[\textbf{(2) }] possible social confusion if subjects later discussed fictional messages that used a coworker name, and
    \item[\textbf{(3)}] psychological confusion or misremembering a simulated message as real.
\end{description}

We mitigated these risks by: 
\begin{description} 
    \item[\textbf{(i)}] containing all messages within the survey (no emails were sent to subjects),
    \item[\textbf{(ii) }] presenting clear consent language that messages were fictional and AI-generated, and
    \item[\textbf{(iii)}] removing identifiers and workplace-context fields from the public research artifact and retaining identifiable source records only in restricted research storage under the approved retention schedule.
\end{description}

Digital records were stored securely in Qualtrics and a restricted university Box environment with research team-only access. Data collection ended on May 31, 2025. Restricted study records held by the research team that are not part of the deidentified public artifact are scheduled for deletion by May 31, 2028, consistent with the approved three-year retention period.

\paragraph{Third-party processing and model use.}
Third-party services included Prolific (for recruitment/payment), Qualtrics (for survey delivery), and an OpenAI model used during the survey to generate simulated spear phishing messages from the provided subject information. During data collection, client-side Qualtrics code transmitted the level-specific workplace fields directly to the OpenAI API. The consent materials disclosed GPT-4o use and represented the model as accessed through the vendor's API policy. The research team did not opt in to OpenAI API data use for model training, and the historical API key has since been revoked. We report this conservatively: OpenAI's API data controls state that API inputs and outputs are not used to train models unless the customer opts in, but default abuse-monitoring logs may retain content for up to 30 days unless an organization has modified abuse monitoring or zero data retention \cite{openai_api_data_controls_2026}. We therefore do not claim that prompts and outputs were never retained by the provider. Generated outputs were stored with the survey records and inspected informally after data collection; participants were not protected by a documented real-time content-screening step.

\paragraph{Deception and debriefing.}
Subjects were told that the messages were AI-generated and simulated. Since messages were not delivered to real inboxes, we reduce the risk of deception-related harm. Due to these conditions, no post-study debrief was needed.

\paragraph{Ethics decision and dual-use considerations.}
Phishing experiments require balancing methodological value against deception, consent, legal, and dual-use concerns \cite{thomopoulos_methodologies_2023}. Because our findings describe what makes phishing pretexts more convincing in a survey setting, they could be misused as well as used defensively. To reduce this risk, we emphasize defensive takeaways (e.g., verification and reporting practices) and we do not release materials that would make it easier to run real attacks, such as subject data, raw message text tied to real organizations, or prompts that are ready-to-use. For transparency and reproducibility, we provide the placeholder-based study prompt in Appendix~\ref{sec:app_prompt}. We do not release participant data, raw messages tied to real workplaces, or tools for delivering operational phishing campaigns. We believe the remaining risk is justified by the benefit of helping defenders understand how personalized pretext cues affect perceived convincingness and reporting, which can improve cybersecurity measures.

\section{Open Science}
We will release the analysis code, deidentified quantitative message-evaluation data needed for the primary personalization and descriptive sender/pretext analyses, sanitized qualitative code and valence labels for aggregate summaries, and deidentified study materials such as the survey instrument, codebook, and consent form with direct contact details removed. The public artifacts contain no source identifiers, participant or coworker names, workplaces, project descriptions, raw generated messages, open-ended response text, demographics, or phishing-background variables, and no reidentification key is retained. Analyses that require restricted demographics or raw qualitative text cannot be reproduced from the public data. A repository link will be provided upon publication.


\section{LLM Prompt}
\label{sec:app_prompt}
Below is the prompt used to generate spear phishing email examples.  
This version corresponds to personalization Level~4. Level~3 omitted COWORKER and PROJECT. Level~2 also omitted JOBRESP1 and JOBRESP2 but retained NAME and JOB\_TITLE. Level~1 retained only WORKPLACE.
The survey made four API calls per participant and requested two outputs in each call. The historical generation code did not retry when an output was not returned; four fallback evaluations were therefore excluded from analysis. Requests used the OpenAI Chat Completions API with gpt-4o, temperature 0.8, top-p 1, and a maximum of 1,000 output tokens. The level-specific prompt was sent as the system message; settings not specified in the deployed code used API defaults.

\begin{Verbatim}[
  breaklines,
  breakanywhere,
  breaksymbolleft={}
]
You are a cybersecurity professor educating your students by simulating a spear phishing attack. 
Your task is to generate two spear phishing emails directed to the fictional user described in their profile below. 

Format each message like a short email.  
Make the two messages as distinct as possible from one another.  
Include a link in each message that appears as '[LINK]'. Never use square brackets anywhere else.  

If you reference people other than those in the profile, use believable names that would plausibly exist in the workplace.  
If you reference another entity or department, it must either appear in the profile or be a widely recognized, real organization.  
Do not invent titles, departments, or institutions. If you are unsure whether one exists, exclude it.  

Each message must include the workplace, the user's name, job title, one job responsibility, the coworker, 
and the project. Be creative in how you weave this information in.  
Use the first job responsibility in the first message, and the second job responsibility in the second message.  

If the message is not official communication, use a more casual tone.  
Keep each message under 150 words, and make them engaging.  

Here is the user's workplace: WORKPLACE  
Here is the user's name: NAME  
Here is the user's job title: JOB_TITLE  
Here is the user's first job responsibility: JOBRESP1  
Here is the user's second job responsibility: JOBRESP2  
Here is a coworker's name: COWORKER  
Here is something they are working on: PROJECT  

Enumerate the messages as 1) and 2).
\end{Verbatim}

\section{Historical Personalization Examples}
\label{sec:app_historical_examples}

The following are altered and deidentified historical outputs shown to one study participant. Identifying profile details, including names, organization, and workplace context, have been replaced with fictional values. The replacements preserve the pretext structure relevant to the comparison but are not verbatim. GPT-4o generated each message independently, so scenario, sender, tone, and request were not held constant across levels. The observed convincingness rating and open-link selection are included only to illustrate that greater information availability did not produce a monotonic response for this participant. Eight additional deidentified historical examples from other participants appear in the reproducibility artifact.

\subsection*{Level~1 (Style~1; rating 100; selected open-link)}

\textit{Available context: workplace.}

\begin{quote}\small
\textbf{Subject: Important: Meridian Mobility Employee Benefits Portal Update}

Hello Team Member,

Exciting news! Meridian Mobility is rolling out new enhancements to our Employee Benefits Portal, designed to provide you with a more seamless experience. To explore these new features and confirm your current benefit selections, please access the portal through the link below:

\texttt{[LINK]}

We encourage you to take advantage of the latest updates tailored to meet your needs. If you have any questions, feel free to reach out to our Benefits Support Team.

Thank you for your continued dedication to Meridian Mobility.

Warm regards,\\
Benefits Administration\\
Meridian Mobility
\end{quote}

\subsection*{Level~2 (Style~2; rating 90; selected open-link)}

\textit{Available context: workplace, name, and job title.}

\begin{quote}\small
\textbf{Subject: Important Change in Production Protocols}

Hi Alex,

I hope this message finds you well. As the Production Operations Manager, it is important for you to be aware of recent updates to our production protocols. We have made several changes to improve efficiency and safety standards. To review these updates, please follow this link: \texttt{[LINK]}. Please familiarize yourself with the new procedures by the end of this week to ensure a smooth transition.

Best regards,\\
Michael Thompson\\
Quality Assurance Team, Meridian Mobility
\end{quote}

\subsection*{Level~3 (Style~2; rating 50; did not select open-link)}

\textit{Available context: workplace, name, job title, and job responsibilities.}

\begin{quote}\small
\textbf{Subject: Important Update Required---Quota Tracking System}

Hi Alex,

I hope this message finds you well. We have recently upgraded our quota tracking system to help you manage your team of 1,200 production staff and meet its targets more efficiently. As the Production Operations Manager at Meridian Mobility, your role is crucial in ensuring smooth operations, and this new tool is designed to support that.

Please review the new features by clicking the following link: \texttt{[LINK]}. Your feedback is invaluable to us, and we believe this update will improve your team's productivity.

Best regards,\\
Jessica Thompson\\
Meridian Mobility IT Support
\end{quote}

\subsection*{Level~4 (Style~3; rating 70; did not select open-link)}

\textit{Available context: workplace, name, job title, job responsibilities, coworker, and project.}

\begin{quote}\small
\textbf{Subject: Development on Automated Line Integration}

Hey Alex,

I wanted to keep you updated on the progress Jordan and I are making with the automated line loaders. With your experience keeping automated assembly equipment running smoothly, we would value your input on integrating the new systems. We are considering several options to optimize the process. Please review our implementation plan and provide feedback here: \texttt{[LINK]}.

Let us know what you think.

Cheers,\\
Morgan
\end{quote}

\bibliographystyle{IEEEtran}
\bibliography{main}

\section*{LLM Usage Statement}
GPT-4o was used as part of the study methodology to generate simulated spear phishing messages from participant-provided work-context fields, as described in the methodology and ethical considerations sections. The authors remain responsible for all text, coding decisions, tables, figures, analyses, and claims.

\end{document}